# Intelligent mode-locked NPR fiber laser based on laser speckle characteristics

Yongjie Pu,[a] Minyu Fan,[a] Zhicheng Zhang,[a] Jie Zhu,[a] Huinan Li,[a] Sha Wang[a,*]

[a] College of Electronics and Information Engineering, Sichuan University, Chengdu 610064, China

**Abstract**：Passively mode-locked fiber lasers based on nonlinear polarization rotation (NPR) have been widely used due to their ability to produce short pulses with high peak power and broad spectrum. Nevertheless, environmental disturbances can disrupt the mode-locked state, making it a challenge for practical implementation. Therefore, scientists have proposed mode-locked NPR lasers assisted with artificial intelligence, which can effectively address the issues related to mode-locking stability. Speckle patterns containing spectral information can be generated when the laser transmitting through a scattering medium, which can be served as indicators of the mode-locked state. The contrast of the Tamura texture feature of the speckle patterns exhibits periodic "V" shaped variations with respect to the rotation angles of the waveplates, according to experimental results. The stable mode-locking region is confined to the area close to the minimum contrast. Based on these characteristics, an intelligent approach employing a modified gradient algorithm to identify the region of minimum speckle contrast for achieving mode-locked state. The average number of iterations needed to achieve initial mode-locking and recover mode-locking are about 20 and 10, respectively. Once the mode-locking is achieved, the neural network can be employed to distinguish single-pulse or multi-pulses outputs based on the speckle pattern, thereby enabling intelligent stable mode-locked single-pulse genration from the NPR fiber laser.

**Keywords:** NPR fiber laser; Speckle; AML; Neural network

* Sha Wang, E-mail: shawang@scu.edu.cn

## 1. Introduction

In recent years, ultrafast fiber lasers have been widely used in various fields such as high-resolution atomic clocks [1-2], optical frequency measurement [3], precise distance measurement [4], signal processing [5], and astronomy [6], owing to their small size, high gain, wide gain bandwidth, and superior beam quality. Mode-locked technology is a key approach for generating optical pulses at the picosecond and femtosecond levels. Passive mode-locked fiber lasers are research hotspots for their superior performance and straightforward structure. The technique based on polarization control and Kerr nonlinear non-polarization rotation are now the preferred method for constructing passively mode-locked lasers. The polarization states can be changed with the light intensity based on Kerr nonlinearity and intensity-dependent nonlinear phase shift in the fiber in the NPR fiber lasers. It is possible to generate efficient artificial saturable absorption by modifying the polarization state inside the cavity. However, environmental factors may cause drifts to the laser polarization state in the fiber [10-11], which leads to fluctuations in the laser mode-locked state. Consequently, automatic mode-locking (AML) technology has emerged as a novel research area for ultrafast lasers and has attracted significant attentions and investigations [12].

AML technology is an endeavor to address the quandary of NPR mode-locked fiber lasers by employing automatic control and various algorithms to adjust specific parameters inside the cavity[13], which establishes a mapping relationship between the polarization state and system parameters. It mitigates polarization state drift brought about by manually controlling or enviromental influence, improving the stability of mode-locked lasers. However, the two urgent progress that should be made for AML technology research are better assessment criteria and faster algorithm. Mode-locked criteria for assessment is typically established based on time-domain waveform and spectral information. In the time domain, the method of pulse counting is generally utilized. X. Shen *et al.* [14] first adopted the technique to achieve AML in a NPR laser. The light extracted from the NPR cavity was converted into an electrical signal through a detector and then was transformed into digital signal after pulse shaping. The number of pulses was counted and the results were transmitted to a microcontroller unit to assess mode-locked state. In 2014, S. Li *et al.* [15] achieved AML within 90 s using pulse counting with a single chip microcomputer (MSP430), two electromagnet polarization controllers, and digital-to-analog converters. In 2017, X. Shen *et al.* [16] made enhancements to the number of counters in the AML laser in order to achieve a higher level of precision in the repetition frequency, thus enabling accurate differentiation between the fundamental frequency mode-locked state and other states. In 2018, G. Q. Pu et al. [17] presented a structure with a simplified feedback device and replaced the pulse shaping circuit with an analog-to-digital converter to achieve mode-locked state by counting the pulses. A shortcut library was established to record the empirical driving voltage of each pulse state for facilitate quick achievement of the mode-locked state. In 2020, G. Q. Pu *et al.* [18] achieved mode-locking with an average time of 1.08 s using the dual-region counting scheme. Utilizing spectral data as a decision criterion for AML is an alternate strategy. In 2017, the correlation coefficient $R^2$ was employed by D. G. Winters et al. [19] as a fitness function between the measured and target spectra to reveal their similarity and the mode-locked state could be verified within a time frame of 90 s. However, utilizing a spectrum analyzer to acquire spectral data take a lot of time. Therefore, in 2020, G. Q. Pu *et al.* [20] employed time stretching technology for real-time collection and analysis of spectral information to assist mode-locked state. Nonetheless, this process necessitates the use of high-precision oscilloscopes, which raised the cost of experiments and created substantial portability challenges.

The feedback algorithm in automatic mode-locking primarily consists of traversal algorithm [14-17] and optimization algorithms [18-20]. Due to its simplicity and convenience, the traversal algorithm was widely used in the earlier age of automatic mode-locking technique. This algorithm is capable of iterating through all polarization states, enabling the verification of the scheme and facilitating the observation of the correlation between the polarization state and the mode-locked state. In 2018, H. Wu et al. [21] successfully achieved automatic search of noise-like pulse and fundamental frequency mode-locked pulse states through the traversal algorithm, which controlled by electric polarization controllers (EPC). In the same year, G. Q. Pu et al. [17] proposed a simplified traversal algorithm to quickly reach a mode-locked state by creating a shortcut library while exploring the enormous polarization space. However, the efficiency of automatic mode-locked lasers using the traversal algorithm is low. As a result, optimization algorithms have become one of the most popular techniques in the area of automatic mode-locked fiber lasers. In 2016, R. I. Woodward et al. [22] utilized a genetic algorithm to optimize the driving voltage of EPC and automatic mode-locking was achieved. In 2020, G. Q. Pu et al. [18] presented an improved genetic algorithm to quickly achieve mode-locked state by skipping the remaining generations when detecting the desired state according to identification criteria.

In this paper, we propose a method to achieve mode-locked state with the speckle characteristics of the output laser passing through a ground glass as the criterion for judgment. The output laser of a NPR laser transmits through a scattering medium and is received by a CCD, which captures the speckle patterns. The Tamura texture feature theory is then used to calculate the speckle contrast [23]. The speckle contrast periodically presents a "V" shape with the rotation angles of the wave plates, where the region of minimum contrast corresponds to the mode-locked state. A gradient algorithm in a feedback loop is adopted to quickly achieve mode-locking by finding the minimum value of the speckle contrast. On average, around 20 iterations are required to initially achieve mode-locked state, while 10 iterations are required for recovering mode-locked state. Furthermore, after automatic mode-locked state is achieved, the neural network is utilized to distinguish the single-pulse or multi-pulses outputs from the NPR fiber laser. The proposed artificial mode-locked method dependes on Tamura textrure feature contrast of the laser speckle patterns is both straightforward and inexpensive, with fewer iterations for both initial achieve mode-locked and recovery mode-locked state.

## 2. Methods and theories

Speckle is an optical phenomenon that arises when coherent light interacts with a rough surface or passes through a medium containing random refractive index fluctuations and an interference pattern comprising bright and dark spots can be formed for the interaction[24]. The transfer matrix (TM) is a useful tool for describing the speckle phenomenon, which characterizes the propagation of light waves in the disordered media. When a light source with a spectral bandwidth of $\Delta\lambda$ illuminates the scattering medium, distinct speckle patterns can be generated by different spectral components. The overall speckle pattern can be deemed as an incoherent combination of speckle patterns from various spectral components. The multispectral transmission matrix $B$ is the sum of $N_\lambda$ monochromatic effective TMs, where $N_\lambda$ represents the number of $\Delta\lambda$ separated monochromatic lights. The formula for this matrix can be described as follows [25]:

$$B = \sum_{\lambda_i=1}^{N_\lambda} T(\lambda_i), \qquad (1)$$

According to theoretical analysis, the transmission matrices of different spectral components are different. Therefore, in numerical simulations of speckle patterns, random complex Gaussian matrices were used to replace transmission matrices [26]. The calculation of speckle patterns with bandwidth increasing from 1 nm to 19 nm with a step size of 2 nm was carried out, and the result is shown in Figure 1 (a). It is clear that as the laser bandwidth increases, the light intensity distribution within the speckle pattern becomes more uniform, and the maximum pixel value of the extracted speckle gradually decreases, which is shown in Figure 1(b). The Tamura texture feature theory is a method to describe texture features. This theory suggests that texture can be described in three dimensions: roughness, directionality, and regularity. Therefore, in this paper, the contrast characteristics of speckle fields are analyzed from more dimensions based on the Tamura texture feature theory, and the contrast definition of speckle images is shown in equation (2) [23]:

$$F_C = \frac{\sigma^2}{\sqrt[4]{\mu_4}} \qquad (2)$$

where $F_c$ describes the brightness of speckle patterns, $\sigma$ is the standard deviation of image gray level, and $\mu_4$ is the fourth-order moment of image gray level.

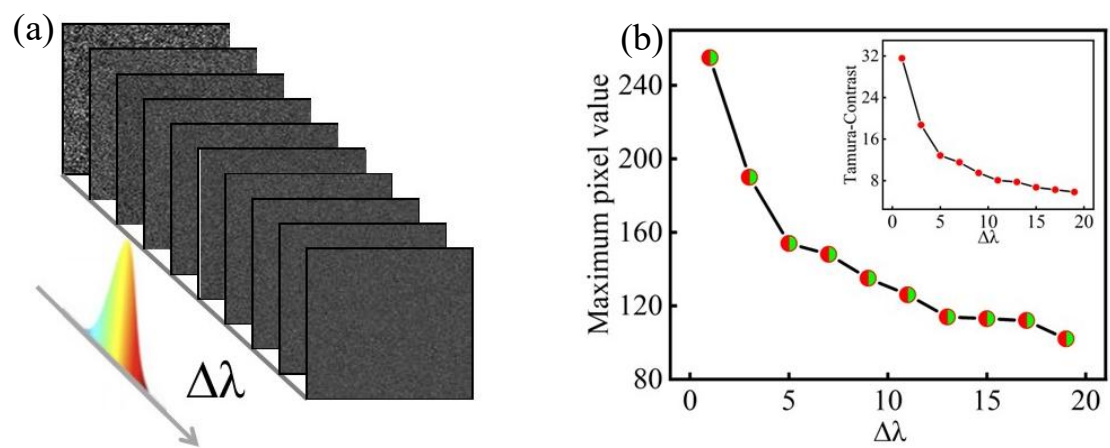

Figure 1. Simulated speckle patterns for different bandwidths. (a) is the speckle pattern corresponding to the beam with gradually increasing $\Delta\lambda$, (b) is the curve graph of the maximum pixel value varies with the bandwidth, and the inset is the curve of the contrast varies with the bandwidth.

Figure 2, shows the change of the Tamura texture feature speckle contrast with the incresing of rotation angle of half waveplate (HWP1) and quarter waveplate (QWP1) in the NPR fiber laser. The curve shows a periodic "V" shape, and the small region near its minimum value indicates the position of the mode-locked state. Therefore, using gradient algorithm to find the minimum value, automatic mode-locking can be quickly achieved.

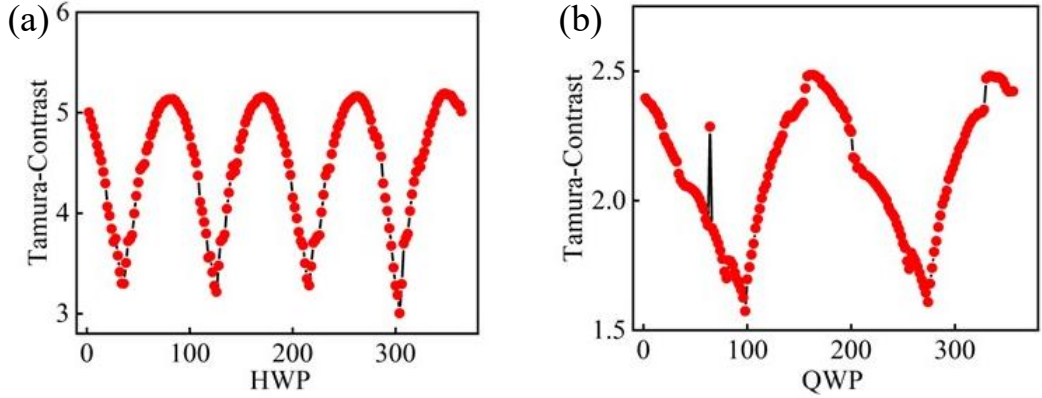

Figure 2. Contrast of speckle patterns generated by a half-wave and a quarter-wave plates after one revolution. (a) Contrast calculation of the speckle pattern induced by a rotating the half-wave plate by applying Equation (2). (b) Contrast calculation of the speckle pattern induced by a rotating the quarter-wave plate by applying Equation (2).

The gradient algorithm [28] is used as an optimization technique for locating the extrema of a function that calculating the gradient of the loss function for each parameter. Then, it updates the parameters in the direction corresponding to the negative gradient until convergence or predetermined conditions are satisfied. In this paper, enhanced gradient algorithm is adopted to optimize the parameters required for achieving AML. The entire process is illustrated in Figure 3. Step 1: Randomly selecting the initial scale X of the half-wave plate, follow by computing the scales of X and X+2° using formula (2) to determine the contrast of the speckle pattern. Finally, the gradient is computed. Step 2: Determining the sign of the gradient ($d_n$). If $d_1>0$, the half-wave plate is rotated counterclockwise by the initial step size of 10° and subsequently positioned at X-10°. Conversely, if $d_1<0$, the wave plate is rotated clockwise using the initial step size of 10° and positioned at X+10°. Then the first step will be repeated to calculate the X-10° and X-10°+2° scales to obtain the contrast and gradient of the speckle pattern. If $d_2>0$, the wave plate is rotated counterclockwise once more using the initial step size of 10°, resulting in it being positioned at X-10°-10°. If $d_2<0$, the wave plate is rotated clockwise by another half step size, which will position it at X-10°+5°. Repeat these steps until the final step length control ceases operation within a tolerance of 2.5°. The same algorithm applies to the quarter-wave plate, except that the initial step size is set to 20°.

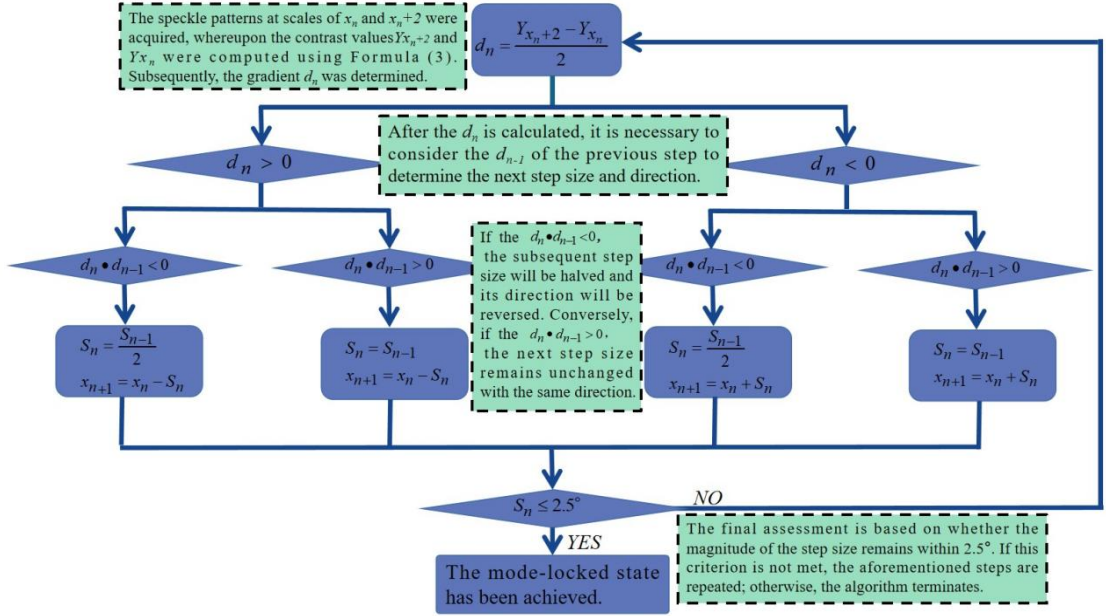

Figure 3. Flowchart for AML using the gradient algorithm. *X* is the scale of the half-wave plate, $d_n$ is the gradient, *Y* is the contrast, *n* is the number of iterations, and $S_n$ is the step size.

The process of laser mode-lockedwith the NPR technique can have an effect similar to saturable absorption, which leads to variations in the spectrum of the resultant output laser, resulting in different speckle patterns or intensities. In order to distinguish the single-pulse or multi-pulse states based on these speckle patterns, a Convolutional Neural Network (CNN) model was employed [40]. Specifically, CNN is trained to classify different mode-locked states based on the corresponding speckle pattern features. The network architecture comprises an input layer, a convolutional layer, a pooling layer, a fully connected layer, and an output layer [29]. The convolutional layer employs a 5×5 convolution kernel to perform convolution operations on the feature map measuring 128×128 pixels. This results in six convolutional feature maps with a size of 124 × 124 pixels. Subsequently, the Rectified Linear Units (ReLU) activation function is introduced for nonlinear computations. The maximum pooling layer is selected to enhance the details for improved recognition. After the pooling layer, the image is transformed into a pooled feature map consisting of 6 pooled feature maps with a size of 62×62 pixels. The fully connected layer consolidates the features extracted through the stacked convolutional and pooling layers, converting the image features into a size of 1×23064 pixels. The CrossEntropy loss function is adopted, and the Adam optimizer is used to train the network. Finally, the speckle pattern is classified as single-soliton speckle, multi-soliton speckle, and speckle in the unlocked-state. Figure 4 presents the network model utilized in this study.

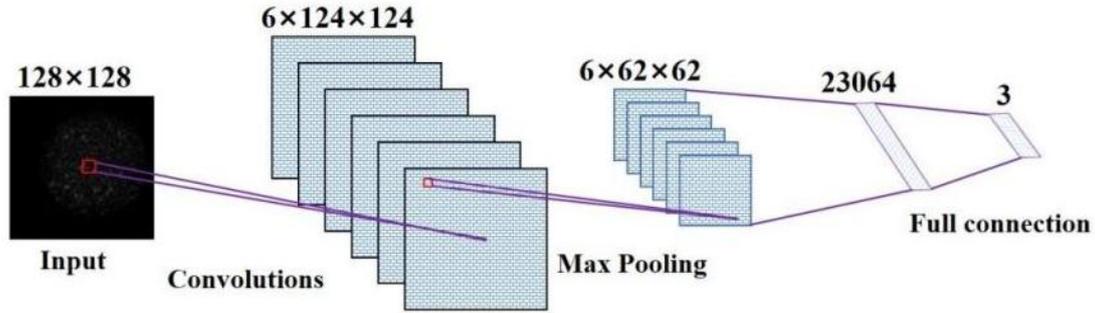

Figure 4. CNN network model.

## 3. Experimental setup

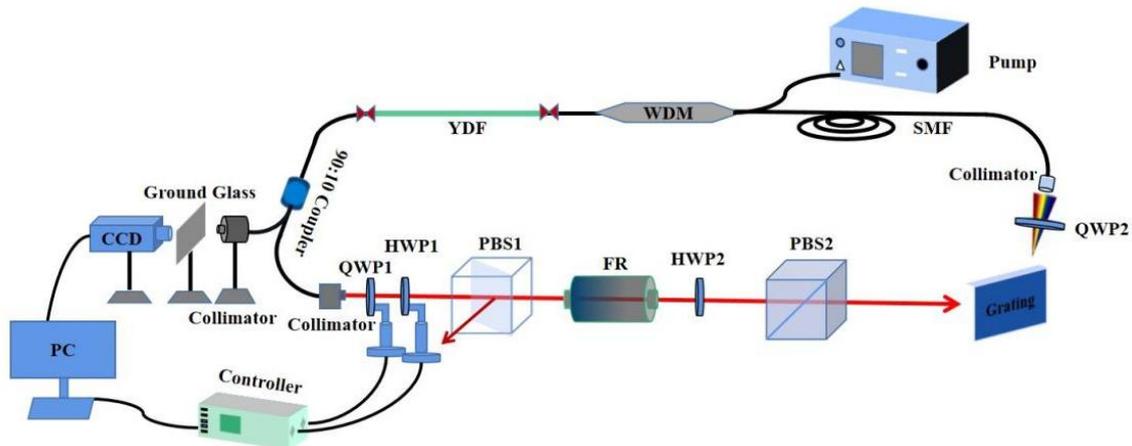

Figure 5. NPR mode-locked fiber laser. Pump: pumping source; WDM: wavelength division multiplexer; SMF: single-mode fiber; YDF: Yb-doped fiber; PBS1, PBS2: polarization beam splitter; Grating: grating; FR: Faraday magneto-optical rotator; HWP1, HWP2: half-wave plate; QWP1, QWP2: quarter-wave plate.

The mode-locked NPR fibre laser in the experiment is shown in Figure 5. The length of the fibre laser is about 2.5 m. A piece of 30 cm long Yb doped fiber (LEIKKI 1200-4/125) was used as the gain medium. The space optical path is about 0.5 m, the net dispersion is about 0.05 ps$^2$, which corresponds to total positive dispersion. A pump source with a pump wavelength of 976 nm, and coupled into the cavity through a wavelength division multiplexer(WDM). Fiber collimators are used to transmit light within an optical fiber into a spatial optical path. A 10:90 beam splitter is used to split the light path into two paths, with 90% of the light entering the spatial light path. Quarter-waveplates, half-waveplates, and a polarization beam splitting prism (PBS) are used to achieve mode-locked state by controlling nonlinear polarization evolution in the cavity. Faraday rotator, a half-wave plate, and longitudinally positioned polarization beam splitter prism form a spatial optical isolator. In order to achieve stable mode-locking under normal dispersion, a 600 lines/mm reflection grating and a collimator are used to form a narrow-band Gaussian spectral filter. The central wavelength can be tuned by controling the filter. The remaining 10% of the light is placed in front of the scattering medium (frosted glass), while the CCD (Daheng Optics, MER-132-43U3M, 1292 * 964) is connected to the back of the scattering medium for speckle generation and detection. A computer serves as the computing center for processing the speckle and running the algorithm, wherein the utilized computer has an i9-10900 CPU, an NVIDIA GeForce RTX 3070 GPU, and 64GB RAM. The K-Cube piezoelectric inertial motor controller (KIM101) controls the ORIC rotary displacement platform (PDR1/M) to rotate HWP1 and QWP1 at a speed of 100°/s. The output spectrum is measured by an optical spectrum analyzer (HEWLETT PACKARD™, Model 70004A). The pulse duration is measured by an autocorrelator FR-103WS.

## 4. Results and Analysis

Figure 6 illustrates the spectra and speckle diagrams for unmode-locked, single-pulse, and multi-pulse scenarios when HWP1 is adjusted at a pump power of 477.8 mW. Figures 6(a) and (b) show the spectra and speckle patterns when the laser is not mode-locking, with a speckle contrast of 8.5943. Figures 6(c) and (d) depict the spectrum and speckle pattern of a single pulse, respectively, with a full width at half maximum of 31.87 nm and central wavelength of 1030 nm. The speckle contrast is 4.6561, and the autocorrelation plot of a single pulse with a full width at half maximum of 2.038 ps is illustrated in the inset in Figure 6(c). Figures 6(e) and (f) illustrate the spectra and speckle patterns of the 3-soliton bound state (identified from the inset of 6(e)), with a full width at half maximum of 25.27 nm. The speckle contrast is 7.4384 and it decreases as the spectral width increases. Therefore, speckle contrast changes can be used to determine the mode-locked state.

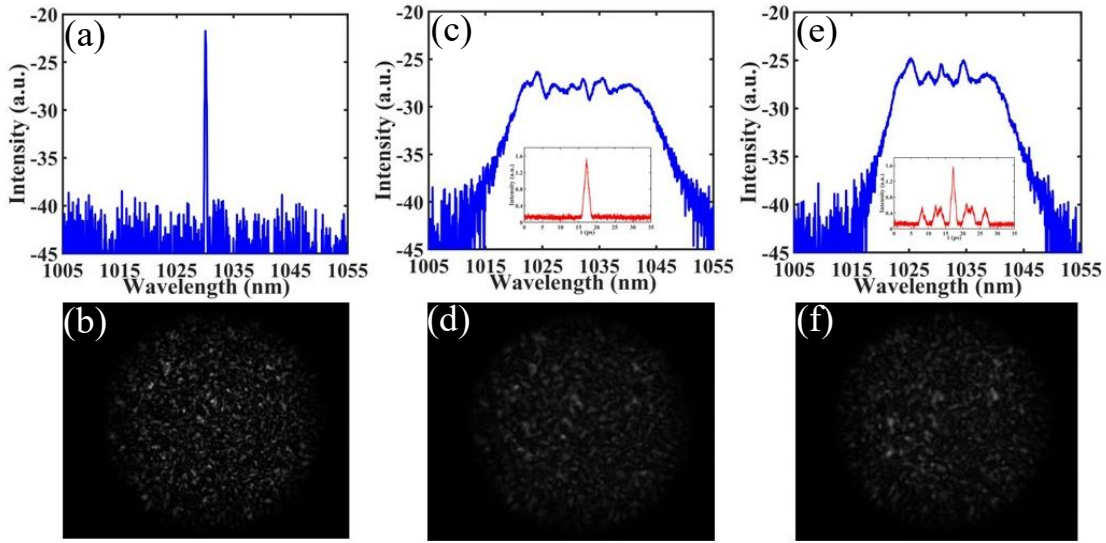

Figure. 6. Spectrum and speckle pattern obtained under unlocked-mode, single-pulse, and multi-pulse states when HWP1 is adjusted. (a) and (b) are the spectrum and speckle patterns of the unlocked-state. (c) and (d) are the spectra and speckle patterns of a single-pulse, the inset is the autocorrelation image, (e) and (f) are the spectra and speckle patterns of the 3-soliton bound state, and the inset is the autocorrelation image.

The duration required for mode-locking achieving in each instance is contingent upon the initial settings of both HWP1 and QWP1. The scale of HWP1 appears in a cycle of 90° degrees (as demonstrated in Figure 2(a)), and it takes up to 14 iterations to reach the vicinity of the minimum point of speckle contrast. However, the scale of

QWP1 appears in a cycle of 180° (as shown in Figure 2(b)) requiring only up to 14 iterations at most to achieve mode-locked state. In this study, results of AML experiments are conducted and presented graphically in Figure 7, which illustrates the number of iterations required for initial achieve mode-locked and recovery of mode-locked state. The green and blue histograms represent the number of iterations while rotating HWP1 and QWP1 respectively, and the red histogram depicts the total number of iterations required. The rotation of the wave plate is controlled by the ORIC rotary displacement platform corresponding to an ergodic method, determining the exact time is limited. However, using an EPC to manage the polarization state of the device can improve the speed of achieving mode-locked and recovering mode-locked. Theoretically, using the ORIC rotary displacement platform at a speed of 100°/s might enable rapid mode-locked within 5.69 seconds. Whereas, practically, the platform exhibits errors when running at high speeds, thereby limiting its accuracy of controlling the rotation angle. Thus, in this article, the number of mode-locked iterations is employed instead of specifying specific times. Upon computing the number of initial achieve mode-locked attempts ten instances, it is observed that the quickest 12 iterations achieved mode-locked state, with an average of 19.4 iterations necessary for achieving mode-locked state. Regarding the statistics for the number of iterations required for mode-locked recovery after losing lock across ten attempts, it is found that the quickest 4 iterations achieved mode-locked state, and an average of 10 iterations are necessary to re-establish mode-locked state. If EPC is employed to regulate the polarization state, mode-locked state can be achieved in 0.24s on average.

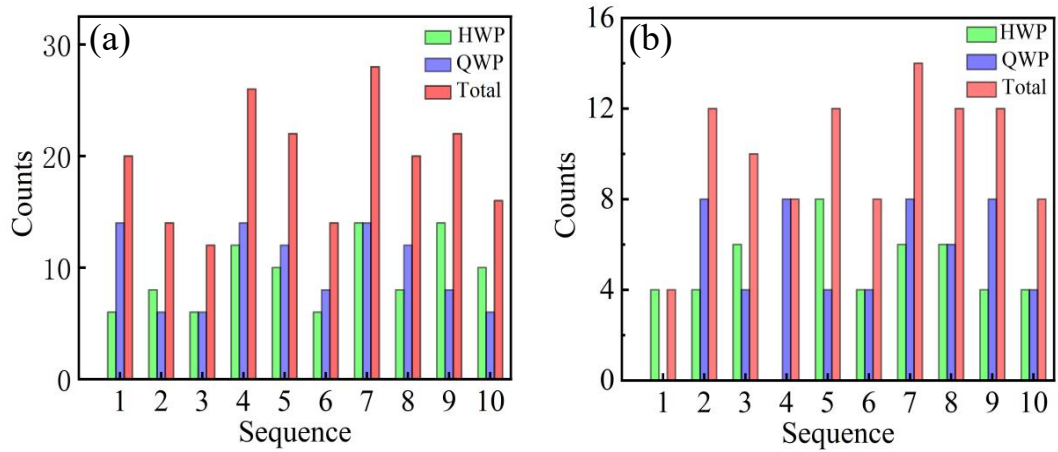

Figure 7. Statistical graph of the number of 10 consecutive AML experiments. (a) The number of initial mode-locked iterations, (b) The number of iterations to recover mode-locked after loss the mode-locked state.

After AML is successfully achieved, a CNN is employed to automatically analyze the current speckle pattern and determine whether it is a single-pulse or multi-pulse speckle pattern. To train the network, 2,000 speckle patterns in different states are collected, including single-pulse, multi-pulse, and unlocked-mode. A total of 6,000 images through translation and rotation methods are generated and then classified into an 80% training set, a 10% validation set, and a 10% test set. Subsequently, the data graphs are fed into the CNN network structure depicted in Figure 4 for training and validation. The loss and accuracy results during iterative training and validation are demonstrated in Figure 8(b) and (c). During the classification stage, owing to the salient features, the CNN is trained rapidly, with the accuracy rate of the training set reaching 100% by the 42 epochs, while the accuracy rate of the validation set reaching 99.7% at the 52 epochs. Finally, the test set is employed to predict the model's performance, with the accuracy rate surpassing 99% after 30 epochs. The results indicate that the network structure can efficiently classify speckle patterns as either single pulse, multi-pulse, or unlocked-mode. If the speckle pattern is identified as mode-locked with multiple pulses, then the laser power can be reduced to achieve single-pulse mode-locked.

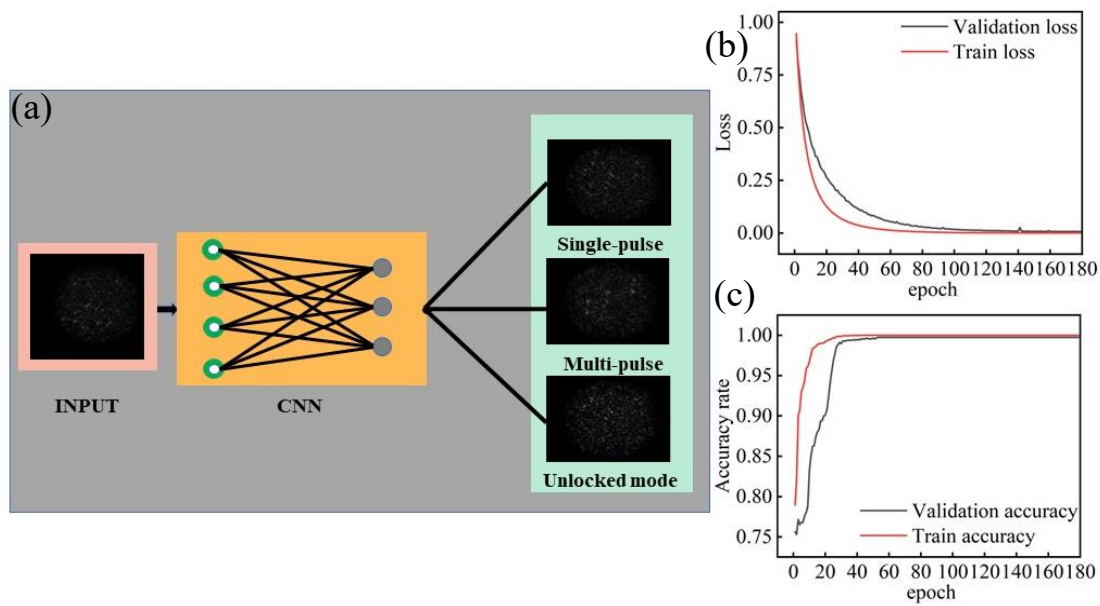

Figure 8. (a) The flow chart of CNN model training, (b) the change curve of loss with the function of epochs in the training process , (c) the change curve of accuracy rate with the function of epochs in the training process.

## 5. Conclusion

In this paper, we present an intelligent mode-locked NPR fiber laser adopting speckle properties characterized by Tamura texture feature theory. By introducing an improved gradient algorithm, the mode-locked region can be rapidly located. The laser has the ability to quickly recover from polarization change due to the environmental disturbances. Further, once the mode-locking is achieved, the neural network is used to determine whether the output laser operates in single-pulse or multi-pulses states according to the speckle pattern. This discovery enables rapid realization of automate mode-locking and replaces conventional spectrometers and autocorrelators for pulse-type evaluation. In future research, the algorithm can be further optimized by simultaneously operating the two waveplates to achieve even less iterations to achive mode-locking. Moreover, we intend to transform our existing NPR fiber laser into an all-fiber structure while employing EPC and a fast frame rate camera which will significantly improve the mode locking speed.

**Funding**

This work was supported by the National Natural Science Foundation of China (61975137).

**Disclosures**
The authors declare that there are no conflicts of interest related to this article.